\documentclass{pasj00}

\begin{document}
\SetRunningHead{T. Sekii et al.}{Hinode/SOT Helioseismology}
\Received{2007/07/27}
\Accepted{}

\title{Initial Helioseismic Observations by Hinode/SOT}

\author{
Takashi \textsc{Sekii}\altaffilmark{1},
Alexander G. \textsc{Kosovichev}\altaffilmark{2},
Junwei \textsc{Zhao}\altaffilmark{2},
Saku \textsc{Tsuneta}\altaffilmark{1},
Hiromoto \textsc{Shibahashi}\altaffilmark{3},
Thomas E. \textsc{Berger}\altaffilmark{7},
Kiyoshi \textsc{Ichimoto}\altaffilmark{1},
Yukio \textsc{Katsukawa}\altaffilmark{1},
Bruce W. \textsc{Lites}\altaffilmark{4},
Shin'ichi \textsc{Nagata}\altaffilmark{5},
Toshifumi \textsc{Shimizu}\altaffilmark{6},
Richard A. \textsc{Shine}\altaffilmark{7},
Yoshinori \textsc{Suematsu}\altaffilmark{1},
Theodore D. \textsc{Tarbell}\altaffilmark{7},
and
Alan M. \textsc{Title}\altaffilmark{7}}


\altaffiltext{1}{National Astronomical Observatory of Japan,
Mitaka, Tokyo 181-8588}
\email{sekii@solar.mtk.nao.ac.jp}

\altaffiltext{2}{W. W. Hansen Experimental Physics Laboratory,
Stanford University, Stanford, CA 94305, USA}

\altaffiltext{3}{Department of Astronomy, School of Science, University of Tokyo, Bunkyo-ku, Tokyo 113-0033}

\altaffiltext{4}{High Altitude Observatory,
National Center for Atmospheric Research, P.O. Box 3000,
Boulder, CO 80307, USA}
\altaffiltext{5}{Kwasan and Hida Observatories,
Kyoto University, Kuravashira, Kamitakara-cho,
Takayama, Gifu, 506-1314}
\altaffiltext{6}{Institute of Space and Astronautical Science,
Japan Aerospace Exploration Agency,
3-1-1 Yoshinodai, Sagamihara, Kanagawa 229-8510}
\altaffiltext{7}{Lockheed Martin Solar and Astrophysics
Laboratory, B/252, 3251 Hanover St., Palo Alto, CA 94304, U.S.A.}


%


\maketitle

\begin{abstract}
Results from initial helioseismic observations by Solar Optical
Telescope onboard Hinode are reported.
It has been demonstrated that intensity oscillation data from
Broadband Filter Imager can be used for various helioseismic
analyses. The $k-\omega$ power spectra, as well as corresponding
time-distance cross-correlation function that promises
high-resolution time-distance analysis below $6$-Mm travelling distance,
were obtained for G-band and Ca\emissiontype{II}-H data.
Subsurface supergranular patterns have been observed from
our first time-distance analysis. The results show that the solar
oscillation spectrum
is extended to much higher frequencies and wavenumbers, and the
time-distance diagram is extended to much shorter travel distances
and times than they were observed before, thus revealing great
potential for high-resolution helioseismic observations from
Hinode.

\end{abstract}

\section{Introduction}

Local helioseismology has been a great success in revealing
subsurface structure of the sun, particularly in detecting
three-dimensional flow and sound-speed anomaly in active regions
(see, e.g., \cite{Zhao_Kosovichev_Duvall_2001},
\cite{Zhao_Kosovichev_2003}).

Observations for local helioseismology require high resolution
and wide coverage both in space and in time.
The 50-cm aperture Solar Optical Telescope
(SOT, \cite{Tsuneta_etal_2007, Suematsu_etal_2007})
onboard Hinode satellite (\cite{Kosugi_etal_2007})
provides $0.2$-arcsec spatial resolution
though with a limited field of view ($328$\,arcsec
$\times 164$\,arcsec for Narrowband Filter Imager (NFI) and
$218$\,arcsec $\times 109$\,arcsec for Broadband Filter
Imager (BFI), nominally).
Except during the satellite eclipse periods,
Hinode can observe the sun continuously with cadence up to $15$\,s
or even less, with the actual limit placed by telemetry bandwidth
and capacity of the data recorder, which are shared with other
instruments as well as other SOT programs running at the same time
frame.

The high resolution of SOT particularly offers a great potential
for local helioseismology.
One of the main scientific targets of Hinode is to investigate
the generation, evolution and eventual dissipation of active regions
on the sun, in attempt to observationally constrain theories
on dynamo processes which are thought to be responsible for
the solar activity cycle.
Local helioseismology can contribute greatly in this area,
not only via active region tomography but also in measuring
flows at larger scales such as meridional flow, which plays
an important role in the current models of solar activity cycle
(\cite{Dikpati_Gilman_2006}).
Please note that even with the limited field of view of SOT,
and hence the limited depth coverage for tomographic study
(see below),
one can still measure near-surface components of large-scale
flows in a small area first but then extend the measurement
to a broader area simply by repeating the observation at
different locations.

It was planned to use Dopplergrams produced by NFI for
helioseismic observations on Hinode
(\cite{Sekii_etal_2001, Sekii_2004}).
Not only oscillation signals in Dopplergrams are known to exhibit
higher signal-to-noise ratio than oscillation signal in
intensity, but also NFI has a larger field of view than BFI.
A larger field of view translates into a better depth coverage
for tomography, because waves that penetrate deeper into the sun also
travels further in horizontal directions.
Thus NFI Doppler observation was the preferred mode of helioseismic
observation by Hinode.

However, long series of Dopplergrams are still only being tested
at the time of writing. Instead, we examined BFI instensity
oscillation data from H line (Ca\emissiontype{II}) and  G band (CH).

The main aim of the present paper is to report that helioseismic
observations can indeed be made with BFI.
As an example, a result from our first attempt at time-distance
analysis of subsurface flow is presented.

\section{Data analysis}
A quiet region around the disk center was observed from around
1600UT, 1 Jan 2007, to around 0400UT, 2 Jan 2007, for about 12
hours with a 12-minute interruption in the middle of the run.
Hinode was tracking the region during the observation (except
during the interruption mentioned in the above) and in addition
Correlation Tracker was used to maximize frame-to-frame correlation
in the central region of the field of view (\cite{Shimizu_etal_2007}).
No further tracking was done on ground,
unlike in the case of a sunspot study (\cite{Nagashima_etal_2007}).

Both G-band images and Ca\emissiontype{II}-H images were acquired at 
approximately 1-min cadence, Ca\emissiontype{II}-H images following G-band images
by $20$\,s.
Sample images are shown in figure 1.
The CCD pixels were $2\times 2$ summed onboard to reduce the
data amount, to double the effective pixel size to $0.106$ arcsec.
We applied another $2\times 2$ summing on ground (the overall
result was a $4\times 4$ summing) for the sake of faster
analyses, though in this case we no longer fully exploit the
$0.2$-arcsec resolution of the telescope.
The benefit of using the full $2\times 2$-summed data,
or observing with the full capacity of SOT (no summing),
will be investigated in future work.

The cadence was not exactly
regular, due to the fact that there are two clocks involved in
controlling the equipment. Not only they are on different
cycles ($0.5$\,s for Mission Data Processor and
$1.6$\,s by Polarization Modulator Unit), the clocks are slowly drifting
away from each other. Although programming tricks to correct this
behavior are being investigated and tested, they have not been
applied to the current datasets.

A simple test was run to check the effect of the slightly irregular
cadence, by putting through a sine wave on this irregular temporal
grid and then computing power spectrum, pretending that the cadence
was exactly 1 minute. As we expected, the effect was insignificant
and therefore we carried out all the following analyses assuming that
the cadence was exactly 1 minute.

The projection effect was also neglected. Because the BFI field of
view is so narrow it should not give rise to any significant
inaccuracy around the disk center. 

We used only the first uninterrupted 6-hr of the data for basic
diagnostics products in the next section. For the time-distance
analysis, the full 12-hr data were used. Before starting the
analyses that follow, first we applied running difference of
images: each image was subtracted from the following image,
thus obviating the need for calibrating each image for dark,
flat field and bad pixels.
One might worry that employing the running difference might
increase the noise level, if the cadence is too short compared
to the time scale over which the signal changes, in which
case the differences between two successive frames are dominated
by noise. This is not the case here, when the 1-min cadence is
not too high for the oscillation signal that is dominated by
5-min oscillations.

\begin{figure}
  \begin{center}
    \FigureFile(80mm,80mm){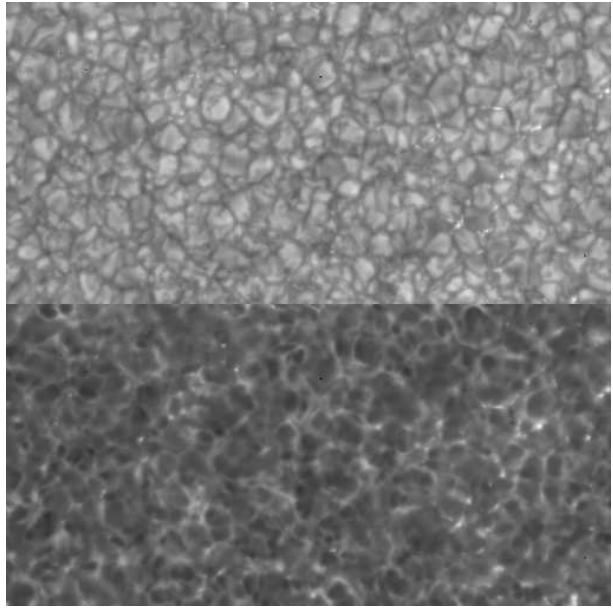}
  \end{center}
  \caption{Sample images of the observed region in G band (top)
and Ca\emissiontype{II} H line (bottom). Of the full field of
view which is $218$\,arcsec $\times 109$\,arcsec, only the central
part of the left half (approximately $54$\,arcsec
$\times 27$\,arcsec) is shown.}
\end{figure}

\section{Power spectra and time-distance diagram}

In this section we present the most basic products of
helioseismic observations: the so-called $k$--$\omega$
diagram (or $l$--$\nu$ diagram) and the time-distance
diagram. They are then compared with results obtained from
Michelson Doppler Imager (MDI) onboard SOlar and
Heliospheric Observatory (\cite{Scherrer_etal_1995}).
The MDI data we used were acquired in the high-resolution
mode. Although they too cover the central region of the disk,
they are not coeval with the SOT data.

Figure 2 shows the $k$--$\omega$ diagrams obtained from G-band and
Ca\emissiontype{II}-H data;
the series of the differential intensity-field data
(intensitygram)
were Fourier transformed and converted to power spectra,
which were then azimuthally integrated in wavenumber
space.
Although no spherical-harmonic decomposition was involved,
degree $l$ was assigned to each component by multiplying
its wavenumber by the solar radius.
The power spectra were corrected for the $\omega^2$ factor,
where $\omega$ is angular frequency,
for we have taken running difference before Fourier
transform.

\begin{figure}
  \begin{center}
    \FigureFile(80mm,80mm){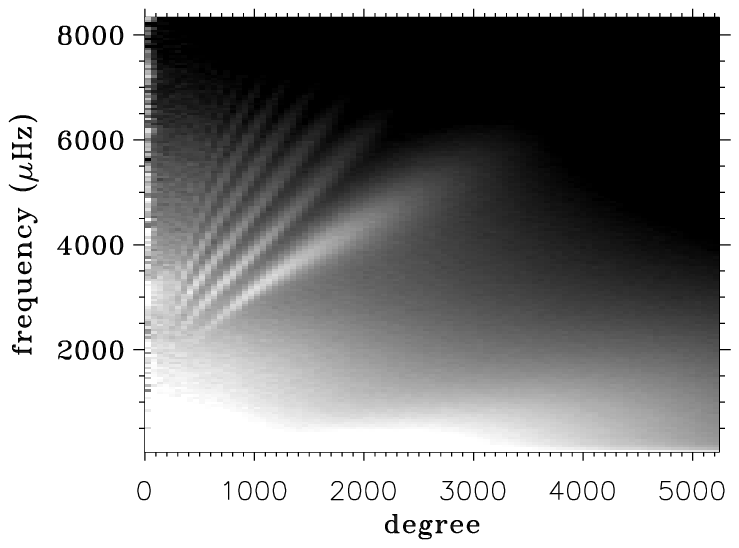}
    \FigureFile(80mm,80mm){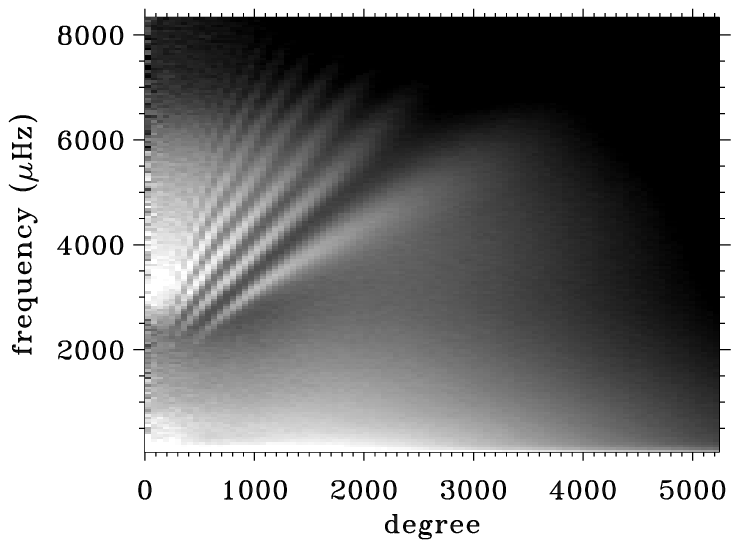}
    \FigureFile(80mm,80mm){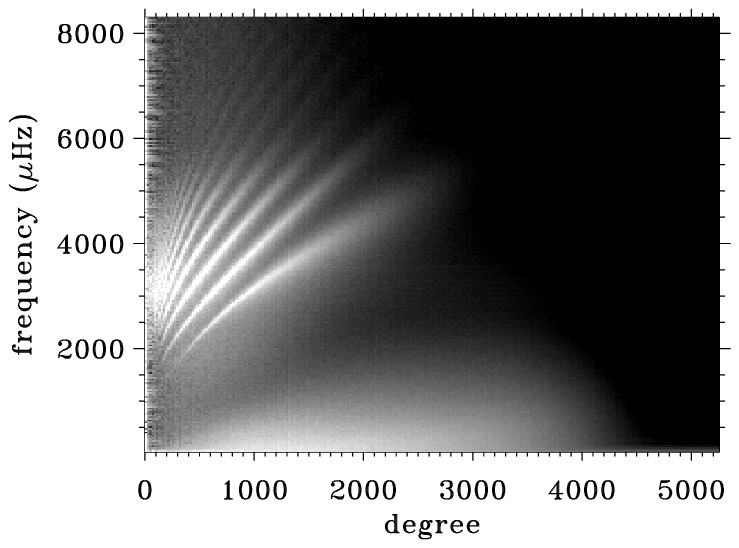}
  \end{center}
  \caption{Power spectra ($k$--$\omega$ diagrams) from the
6-hr G-band data (top) and Ca\emissiontype{II}-H data (middle), compared with
power spectrum from a 512-min MDI high-resolution Doppler data (bottom).}
\end{figure}

In spite of only 6 hours of observation, the f-mode ridge
is seen up to $l\approx 4000$.
We also see the high-frequency interference peaks 
(\cite{Jefferies_etal_1988}, \cite{Duvall_etal_1991},
\cite{Kumar_etal_1990})
up to
the Nyquist frequency. They seem disturbed close to the
Nyquist frequency because the ridges from both the super-
and sub-Nyquist ranges are running into the other regime.
Compared to the $k-\omega$ diagram obtained from a 512-min
MDI high-resolution velocity dataset, also shown in figure 2,
the better performance of SOT at high-wavenumber,
where high-resolution matters most, is noticeable in how
the f-mode ridge extends to higher wavenumbers and,
in the case of Ca\emissiontype{II} H data, the p$_1$ ridge
as well.
However, as was expected, the SOT
$k-\omega$ diagrams show higher background-noise level
compared to the Doppler measurement by MDI.

It is apparent that in G-band data the convective 'noise'
is stronger and for this reason, in the following analyses,
we use only Ca\emissiontype{II} H data. More detailed
comparison between the two, in particular phase relations
at various frequencies, would be a subject of future work.

\begin{figure}
  \begin{center}
    \FigureFile(80mm,80mm){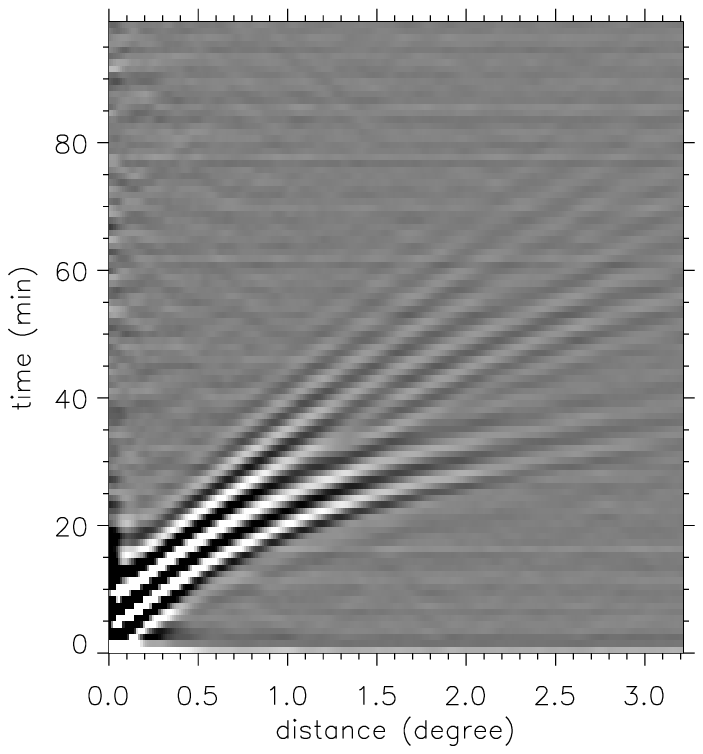}
    \FigureFile(80mm,80mm){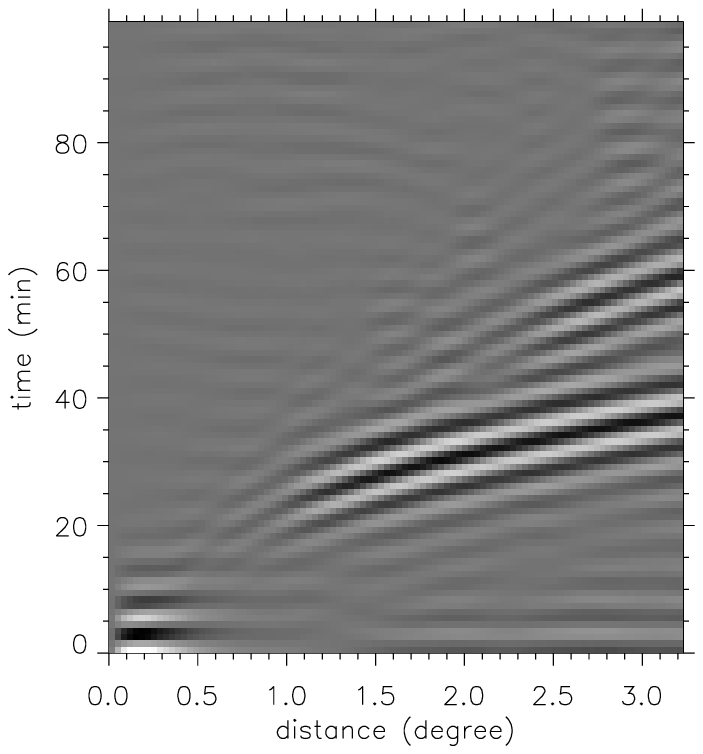}
  \end{center}
  \caption{Time-distance diagrams from 6-hr Ca\emissiontype{II}-H data (top)
and from 512-min MDI high-resolution Doppler data (bottom). The datasets are
not co-eval.}
\label{fig:t-d}
\end{figure}

We then applied Fourier-Bessel transform
(see, e.g., \cite{Sekii_Shibahashi_2003})
to the Ca\emissiontype{II}-H $k$--$\omega$ power spectrum to obtain
time-distance cross-correlation function
(figure \ref{fig:t-d}).
The power spectrum uncorrected for the $\omega^2$ was
used;
this may be viewed as an application of a broad high-pass
filter to the power spectrum,
the intention being filtering out the convective noise.

The structure down to less than $0.1$ heliographical
degree is seen with a high signal-to-noise ratio,
which is very encouraging.
Once again, we see the higher performance of SOT, compared to
MDI, also shown in figure 3, at small spatial scales.
The signal-to-noise ratio, however, is higher for MDI at large
distances and thus the both instruments complement each other.

\section{Subsurface flow}

\begin{figure}
  \begin{center}
    \FigureFile(80mm,80mm){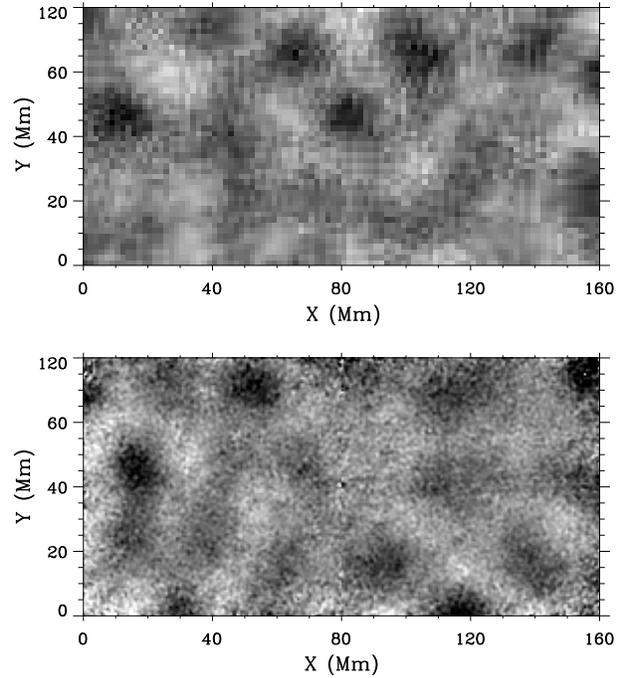}
  \end{center}
  \caption{An outward-inward travel time difference maps
from 512-min MDI high-resolution Doppler data (top)
and from Ca\emissiontype{II}-H data (bottom),
indicating flow divergence due to supergranulation.
The inner radius of the annulus used was $13.86$\,Mm and
the outer radius $15.86$\,Mm.
Dark patches are where inward travel time is longer than
outward travel time i.e. where the flow is generally
diverging, whereas brighter patches are where the flow is generally
converging. Since the two datasets are not coeval, individual
features are not to be compared directly.}
\label{fig:td-flow}
\end{figure}

\begin{figure}
  \begin{center}
    \FigureFile(80mm,80mm){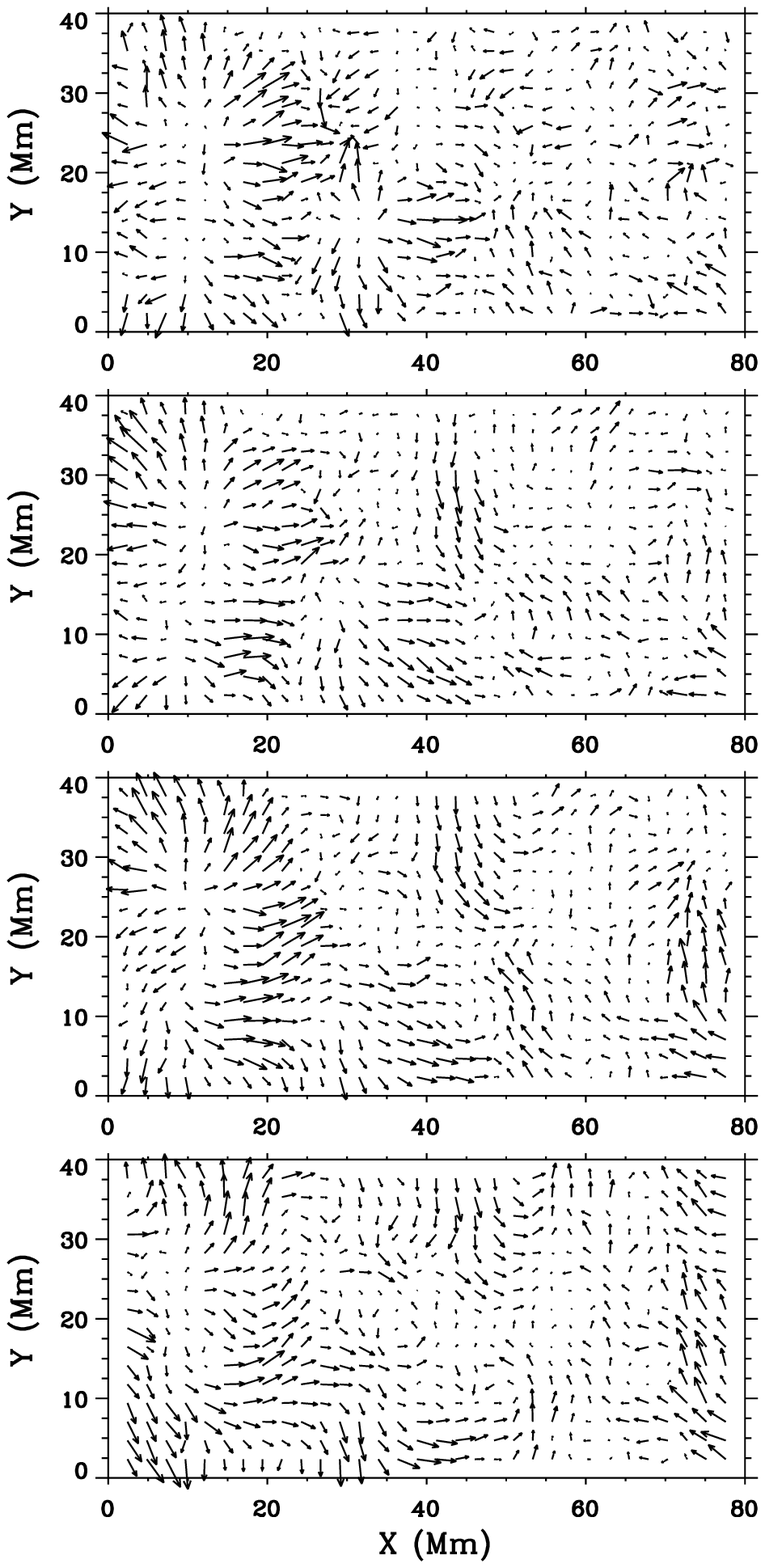}
  \end{center}
  \caption{Subsurface flow maps (indicated by arrows) obtained
by time-distance analysis of 12-hr Ca\emissiontype{II}-H data. Depth ranges are
$0-1$\,Mm, $1-2$\,Mm, $2-3$\,Mm and
$3-4$\,Mm (top to bottom).
The longest arrow in the panels correspond to
$0.51$\,km s$^{-1}$, $0.52$\,km s$^{-1}$, $0.30$\,km s$^{-1}$ and
The field of view corresponds to the central quarter
of the field of view in figure 4.
}
\label{fig:td-flow}
\end{figure}

We have analyzed the full 12-hr long series by time-distance
technique for local helioseismology. The gap between the two
6-hr stretches was filled with zeroes. Then for measuring
travel-time difference between outward and inward wave components,
each pixel was cross-correlated with average signals over annuli
that cover, in radii, 4.00-6.62, 6.47-9.70, 9.24-12.47, 12.01-15.86,
15.09-18.94, 18.48-22.33, and 21.56-25.41 (all in Mm).
One such example as well as an MDI counterpart (in high-resolution
mode) is shown in figure 4.
The annuli were further divided into 4 sectors
representing north, south, east and west, and opposing pairs
of sectors were similarly cross-correlated to yield
northward-southward and westward-eastward travel-time differences.

All the travel-time differences were then inverted by using
ray-approximation kernel for acoustic wave propagating
through the solar interior. Figure 4 shows outward-inward
travel-time difference map, as a proxy divergence of the flow
field. Inversions for horizontal flow
in the depth ranges of 0--1\,Mm, 1--2\,Mm, 2--3\,Mm and 3--4\,Mm
are shown in figure 5.
Inversions were done for the entire area in figure 4,
but to avoid overcrowding the figure with arrows,
in figure 5 only the central quarter is shown.
Horizontal flow patterns are largely consistent with the
outward-inward travel-time difference, as was seen in
realistic numerical simulation by \citet{Zhao_etal_2007}.
This indicates that the travel-time difference is a good
proxy of flow divergence,
though to acquire additional details we do need inversion
procedure. We see that supergranulation patterns
are coherent vertically within a spatial scale of a few Mm.

\section{Discussion}

As we have demonstrated, SOT observation of intensity oscillations
can be used for high-resolution helioseismic diagnosis of the sun,
in spite of its noise level higher than what can be expected from
Doppler measurement. This is due to the fact that solar noise is
higher in intensity than in Doppler velocity.

From figure 2 it is apparent that SOT offers good opportunity
to study high-wavenumber waves, particularly f modes.
\citet{Ishikawa_etal_2007} has found ubiquitous small-scale
horizontal magnetic field from Hinode/SOT observation of the
photosphere. In addition to turbulence, presence of such magnetic
fluxes would affect propagation of f-mode waves and this would
be an important future work.

In the time-distance cross-correlation (figure 3), the structure
below $0.5$ degree, which is about the distance of $6$ Mm, has
been observed much clearer than before. \citet{Sekii_2004}
used G-band data obtained from $50$-cm La Palma SVST to
demonstrate somewhat similar enhancement in spatial resolution
but the noise level was much higher. It was speculated that
the noise came mainly from atmospheric seeing and that the
spaceborne SOT would perform better. Our current results
show that SOT indeed does.
Figure 4 demonstrates that study of fine structures in
supergranulation patterns benefits from the high resolution
of SOT; the bright and dark patches have a typical spatial
scale of $20$\,Mm but a transition from bright to dark
takes place over the scale of less than $10$\,Mm, which
MDI have difficulty in resolving. With SOT results we see
smoother transitions.

To exploit this small-scale information would require more
tuning and testing of the inversion, particularly the setup
of annuli. The trade-off between resolution and error need
to be examined carefully. Once done, we will certainly achieve
higher horizontal resolution in time-distance inversions.
A similar improvement in performance is expected in resolving,
vertically, layers immediately below the surface, since SOT
can use higher-wavenumber and hence shallower-penetrating
waves. This applies also to sound-speed inversion, which
has not been attempted yet.

Time-distance analysis in active regions, be it for flow or
for thermal (and even magnetic) structure, would need further care
particularly if we wish to go a step beyond phenomenological
inversions; wave propagation in magnetized atmosphere has to be
somehow taken into account. For the initial analysis of sunspot
oscillation data, see \citet{Nagashima_etal_2007}.


\bigskip
Hinode is a Japanese mission developed and launched 
by ISAS/JAXA, collaborating with NAOJ as a domestic partner, 
NASA and STFC (UK) as international partners. Scientific operation 
of the Hinode mission is conducted by the Hinode science team 
organized at ISAS/JAXA. This team mainly consists of scientists 
from institutes in the partner countries. 
Support for the post-launch 
operation is provided by JAXA and NAOJ (Japan), 
STFC (U.K.), NASA, ESA, and NSC (Norway).
This work was carried out at the NAOJ Hinode Science Center,
which is supported by the Grant-in-Aid for 
Creative Scientific Research 
``The Basic Study of Space Weather Prediction'' from MEXT, Japan 
(Head Investigator: K. Shibata), generous donations 
from Sun Microsystems, and NAOJ internal funding.


\end{document}